\documentclass[twocolumn,aps,prl,amssymb,amsmath,
 twoside,nofootinbib,preprintnumbers]{revtex4}
\usepackage[T1]{fontenc}
\usepackage[dvips]{graphicx}

\newcommand{\C}{\mathbb{C}}
\newcommand{\R}{\mathbb{R}}
\newcommand{\cH}{\mathcal{H}}
\newcommand{\cS}{\mathcal{S}}
\newcommand{\tens}{\otimes}
\newcommand{\xd}{\mathrm{d}}
\newcommand{\xD}{\mathcal{D}}
\newcommand{\vx}{\roarrow{x}}
\newcommand{\vy}{\roarrow{y}}

\begin{document}
\title{A ``general boundary''
  formulation\\ for quantum mechanics and quantum gravity}
\author{Robert Oeckl}\email{oeckl@cpt.univ-mrs.fr}
\affiliation{Centre de Physique Th\'eorique,
CNRS Luminy, 13288 Marseille, France}
\preprint{CPT-2003/P.4543}
\date{August 13, 2003}

\begin{abstract}
I propose to formalize quantum theories as
topological quantum field theories in a generalized sense,
associating state spaces with boundaries of arbitrary (and possibly
finite) regions of
space-time.
I further propose to obtain such ``general boundary'' quantum
theories through a generalized path integral quantization.
I show
how both, non-relativistic
quantum mechanics and
quantum field theory can be given a ``general boundary''
formulation. Surprisingly, even in the non-relativistic case,
features normally associated with quantum field
theory emerge from consistency conditions. This includes states with
arbitrary particle number and pair creation.
I also note how three dimensional quantum gravity is an
example for a realization of both proposals and suggest to apply them
to four dimensional quantum gravity.
\end{abstract}

\maketitle

One obstacle in the quest for a quantum theory of gravity
appears to be the fact that the foundations of quantum mechanics (QM)
are inherently non-covariant. On the other hand, quantum field
theory (QFT) can be formulated in a covariant way. The
price for this is a globality (manifest in the path integral expression for
$n$-point functions) that fixes space-time to be Minkowskian. Well
known difficulties result from this already for the extension of QFT
to curved space-times.

I propose an approach to formulating quantum theories that is at the
same time local and inherently compatible with special or general
covariance. The main idea is firstly, to associate state spaces with
boundaries of general regions of space-time. Secondly, amplitudes are
determined by a complex function for each region and associated state
space. Crucially, (and contrary to standard QM) connected boundaries
of compact regions are the main focus of attention. In this sense the
formulation is ``holographic'', i.e.\ the information about the
interior of a region is encoded through the states on the boundary.
These structures are required to be coherent in the sense of
topological quantum field theory (TQFT) \cite{Ati:tqft}. This does not
mean that the underlying structure is necessarily topological. For QM
and QFT the relevant background structure is the metric. Only for
quantum gravity would the theory be topological (more precisely
differentiable) in the usual sense. Note that this does not imply a
lack of local degrees of freedom (see \cite{Oe:catandclock}).

Since the association of states with possibly time-like hypersurfaces
is a quite radical step for QM it is crucial to understand their physical
meaning. This is particularly true for particle states in QFT. Thus it
makes sense even with the goal of quantum gravity in mind first to
reformulate non-relativistic quantum mechanics (NRQM) and QFT
in the ``general boundary'' sense. This is the main focus
of this letter.

To achieve this goal I make a second proposal in the form of a
quantization scheme. It turns out that the path integral
approach \cite{Fey:stnrqm} is particularly suitable and my scheme is a
rather straightforward generalization of it.
It is designed so as to produce
``general boundary'' theories of the type described above.
It is ``holographic'' not only in the sense mentioned above but also
in the sense that the underlying classical configuration space on the
boundary should be chosen such that it uniquely encodes a solution of the
equations of motions in the interior. 

I show that both NRQM of particles in a potential as well as scalar
perturbative
QFT are obtainable as holographic quantizations of the relevant
classical theories. This then allows the generalization
to the ``general boundary'' formulation. Surprisingly, in the case of
NRQM features of quantum field theory emerge, such as the necessity
for states of any particle number and the suggestion of pair
creation. Finally, I remark on the fact that quantum gravity
in three dimensions is a TQFT, i.e.\ has a ``general boundary''
description. What is more, it can be obtained precisely as a
holographic quantization and thus fits perfectly into the
scheme. 

Another main motivation for this work comes from an analysis of the
measurement problem in quantum gravity.
As I have argued in \cite{Oe:catandclock}, this suggests precisely a
``general boundary'' formulation for quantum gravity. Interpretational
implications of the ``general boundary'' formulation
are also investigated there.

\section{``General boundary'' formulation}

Let me explain what the ``general boundary'' formulation, i.e.\ a
(suitably adapted notion of) TQFT is.
Let $M$ be a region of space-time (i.e.\ a four-dimensional
manifold) and $S$ its boundary hypersurface. (For the moment I do not
specify what background structure this entails.)
(T1) Associated with each such boundary $S$ is a vector space $\cH_S$
of \emph{states}.
(T2) If $S$ decomposes into disconnected components\footnote{One might
  also want to admit the decomposition of connected boundaries, but
  this will not be discussed here.}
$S=S_1\cup\dots\cup S_n$ then the state space decomposes into a tensor
product $\cH_S=\cH_{S_1}\otimes\cdots\otimes \cH_{S_n}$.
(T3) For a given boundary $S$, changing its orientation, i.e.\ the
side on which it bounds a region $M$, corresponds to replacing $\cH_S$
with the dual space $\cH_S^*$.
(T4) Associated with $M$ is a complex function
$\rho_M:\cH_S\to\C$ which associates an \emph{amplitude} to a state.
One may also
dualize boundaries. This means that one may convert
$\rho_M:\cH_{S_1}\otimes\cdots\otimes \cH_{S_n}\to \C$ to a function
$\rho_M:\cH_{S_1}\otimes\cdots\otimes \cH_{S_k}\to
\cH_{S_{k+1}}^*\otimes\cdots\otimes \cH_{S_n}^*$, replacing spaces
with dual spaces. Mathematically both
versions of $\rho_M$ are equivalent, giving one determines the other
(hence the same notation).
A crucial property is the \emph{composition rule}.
(T5) Let $M_1$ and
$M_2$ be two regions of space-time that share a common boundary
$S$. Let $M_1$ also have a boundary $S_1$ and $M_2$ a boundary
$S_2$. Consider $\rho_{M_1}:\cH_{S_1}\to \cH_{S}$ and
$\rho_{M_2}:\cH_{S}\to \cH_{S_2}$.
(The state spaces are chosen with respect to suitable orientations of
the boundaries.) Then gluing gives $M=M_1\cup M_2$ with boundaries
$S_1$ and $S_2$. The composition rule demands the equality
$\rho_M=\rho_{M_2}\circ\rho_{M_1}$.

For quantum gravity the background structure coming with space-time
regions and their boundaries is just a differentiable structure. One
obtains essentially a proper TQFT. For QM and QFT the
background structure is a fixed metric (usually that inherited from
Minkowski space).

\section{Holographic quantization}

Consider a classical field theory with a given set of fields
$\phi(x)$ and an action $\cS[\phi]$ so that the equations of motion can
be derived from a minimization of $\cS$. (For simplicity I use the
notation of a single scalar field.) Now let $K_S$ be the space of
field configurations on a hypersurface $S$ bounding a region $M$.
The guiding principle is here, that the amount of boundary data
encoded in $K_S$ should be such that it essentially uniquely determines
a classical solution inside $M$ in a generic situation (e.g.\ $M$ a
4-ball). 
(Q1) The space of states $\cH_S$ associated with $S$ is the space of
complex valued functions $C(K_S)$ on $K_S$. This means adopting a
\emph{state functional} picture.
(Q2) The amplitude $\rho_M$ for a state $\psi\in\cH_s$ is given by the
expression
\[
 \rho_M(\psi)=\int_{K_S} \xD\phi_0\, \psi(\phi_0)
 \int_{\phi|_S=\phi_0} \xD\phi\, e^{\frac{i}{\hbar} \cS[\phi]} .
\]
The first integral is over field
configurations $\phi_0$ on $S$. The second integral is over all (not
necessary classical) field configurations $\phi$ inside $M$ that match
the boundary data $\phi_0$ on $S$.

Note that (Q1) gives a prescription for (T1) and ensures (T2) since for
$S=S_1\cup S_2$ a disjoint union, $K_S=K_{S_1}\times K_{S_2}$ and
hence $C(K_S)=C(K_{S_1})\tens C(K_{S_2})$. (T4) is determined by
(Q2). The dualization of boundaries corresponds simply to leaving the
evaluation with a state on those boundaries open. Let $M$
have boundaries $S_1$ and $S_2$ and consider states $\psi_1\in
\cH_{S_1}$ and $\psi_2\in \cH_{S_2}$. Then $\rho_M(\psi_1)$ is an
element of $\cH_{S_2}^*$, i.e.\ a linear map $\cH_{S_2}\to\C$ by
mapping $\psi_2$ to
\[
 \int_{K_{S_1}\times K_{S_2}} \xD\phi_1 \xD\phi_2\,
 \psi_1(\phi_1) \psi_2(\phi_2)
 \int_{\substack{\phi|_{S_1}=\phi_1\\ \phi|_{S_2}=\phi_2}}
  \xD\phi\, e^{\frac{i}{\hbar} \cS[\phi]} .
\]
This also explains (T3).
The composition property (T5) is also rather obvious: Consider an
integral over
all field configurations in two regions with fields fixed on a common
boundary and integrate also over the boundary values. Then this is the
same as doing the unrestricted integral over field configurations in the
union of the regions.

This heuristic quantization procedure based on the path integral thus
leads to ``general boundary'' type quantum theories. The TQFT-like
axioms (T1) - (T5) are automatically satisfied.
I will refer to it as ``holographic quantization'' for the reasons
given above.
The quantization prescription is meant to be neither
precise nor complete. In particular, one would usually ``divide out''
symmetries either from the configuration space $K_S$ or from the
functions $C(K_S)$ on it to arrive at the ``physical'' state space
$\cH_S$. In quantum mechanics an example is the symmetrization or
antisymmetrization for identical particles (see below). In
quantum gravity an important step would be to divide out
diffeomorphism symmetry (see also below).

\section{Quantum mechanics}

Let me show how the standard formalism of quantum mechanics
fits into the ``general boundary'' scheme.
Space-time is now Euclidean or Minkowski space. I denote a point in
space-time by coordinates $(\vx,t)$. The background structure is the
metric which is inherited by regions and their boundaries.
The regions $R$ we admit
are time intervals $[t_1,t_2]$ extended over all of
space. The boundaries $S$ are thus pairs of time-slices $S=S_1\cup S_2$
with $S_1$ at $t_1$ and $S_2$ at $t_2$. According to (T2)
$\cH_S=\cH_{S_1}\tens \cH_{S_2}$ and because of (T3)
$\cH_{S_2}=\cH_{S_1}^*$. Indeed, let $\cH_{S_1}=\cH$ be the
Hilbert space of quantum mechanics and define $\rho_R:\cH\to\cH$ to
be the time-evolution operator $e^{-i/\hbar H (t_2-t_1)}$. Then a state
in $\cH_S$
corresponds to a pair of states $\psi\in\cH$ at time $t_1$ and
$\eta\in\cH^*$ at time $t_2$ (or a linear combination of such
pairs). The transition amplitude between $\psi$ and $\eta$ is
given by $\rho_R$ via
\begin{equation}
 \langle \eta| e^{-\frac{i}{\hbar} H (t_2-t_1)}|\psi\rangle
 = \rho_R(\psi\tens\eta) .
\label{eq:qmtampl}
\end{equation}
The composition property (T5) encodes the composition of time
evolutions. The inverse operation corresponds to the insertion of a
complete set of states $\sum_{\psi} |\psi\rangle\langle\psi|$ at a
given time. 

So far we have only reformulated quantum mechanics. To show that a
generalization to a ``general boundary'' formulation is possible
it is necessary to consider a
specific theory. The idea is to show that this theory arises as a
holographic quantization. Then, the generalization should be
straightforward, being determined by (Q1) and (Q2).

Let me specialize to the NRQM of
scalar particles. Start with just one free particle. The action is
$\cS[\vx]=\int \xd t\, \frac{1}{2} m \dot{\vx}^2(t)$ for a path
$\vx(t)$ and a classical solution
of the equations of motions is a straight line in space-time. It
intersects each time-slice exactly once. Thus for a region $R$
determined by a time interval $[t_1,t_2]$ as above, denote the
intersections with the boundaries $S_1, S_2$ by $\vx_1, \vx_2$. The
configuration space on the boundary $K_S$ associated with $S=S_1\cup
S_2$ which determines a classical solution uniquely is thus the space
of pairs $(\vx_1,\vx_2)$, i.e.\ $\R^3\times \R^3$. According to (Q1) we
should set $\cH_S=C(K_S)=C(\R^3\times \R^3)$. For the disconnected
components we get that $\cH_{S_1}$ and $\cH_{S_2}$ can be identified
with $C(\R^3)$. Indeed, this
is the correct state space for a particle at time $t$, namely the space
of fixed-time wave functions $\psi_{t}(\vx)$. An element
$\Psi(\vx_1,\vx_2)$ of $\cH_S$ is
generally a linear combination of products
$\psi_{t_1}(\vx_1)\eta_{t_2}(\vx_2)$. (Q2) tells us that
$\rho(\psi\tens\eta)$ is given by
\begin{equation}
 \int_{\R^3\times\R^3} \xd \vx_1 \xd \vx_2\, \psi(\vx_1)\eta(\vx_2)
 \int_{\substack{\vx(t_1)=\vx_1\\ \vx(t_2)=\vx_2}} \xD \vx\,
 e^{\frac{i}{\hbar} \cS[\vx]} .
\label{eq:qmpi}
\end{equation}
Here the second integral is over all paths $\vx(t)$ in the interval
$[t_1,t_2]$ with the given boundary values. This is indeed the correct
expression for the transition amplitude (\ref{eq:qmtampl}) in the path
integral formulation of NRQM. Note that we can
easily generalize (\ref{eq:qmpi}) to include a potential in the action.

The extension to several particles is rather obvious. For example, for
two particles, $K_S$ would be the space of
quadruples $(\vx_1,\vy_1,\vx_2,\vy_2)$, while $K_{S_1}$ would be given by
pairs $(\vx_1,\vy_1)$ etc. $\cH_{S_1}$ would be given by $C(K_{S_1})=
C(\R^3\times \R^3)$, i.e.\ fixed-time wave functions $\psi(\vx,\vy)$ of
two particles. (\ref{eq:qmpi}) is generalized in the obvious way
with the path integral now over one path for each particle. This gives
the correct results for \emph{distinguishable} particles. For
\emph{identical} (and bosonic) particles we have to take for $\cH_{S_1}$
the subspace of symmetric functions in $(\vx,\vy)$. A different way to look
at this is to replace the space $K_{S_1}$ of ordered pairs by the
space of unordered pairs. Of course this is not something coming out
of the quantization prescription sketched above, but compatible with
it. Note also that the path integral in (\ref{eq:qmpi}) does not have
to be explicitly symmetrized as it is always evaluated with
symmetrized functions.

The next step is to see how using the rules (Q1) and (Q2) the
NRQM of particles extends to a
``general boundary'' formulation. Consider a 4-ball shaped region $B$
in space-time
with boundary $S$. As with the spatial slices, a classical particle
trajectory intersects $S$ exactly twice. Thus, the configuration space
is essentially $K_S=S\times S$. However, the entry time of the
particle into $B$ is necessarily earlier than the exit time. Thus
$K_S$ is really the subspace of $S\times S$ where one point (say the
first one) has a smaller time coordinate. By (Q1) then $\cH_S$ is the
space of functions $\psi(z_\text{in},z_\text{out})$ on $S\times S$
with this restriction. Here $z$ denotes a parameterization of the
hypersurface $S$. By (Q2) then we have a function $\rho_B$ that
associates amplitudes with such a generalized wave function
$\psi$. The physical interpretation is that of the amplitude of a
particle being sent into the region $B$ at $z_\text{in}$ (which
includes a time coordinate $t_\text{in}$) and being observed emerging
from $B$ at time and place $z_\text{out}$.

\begin{figure}
\includegraphics[scale=0.8]{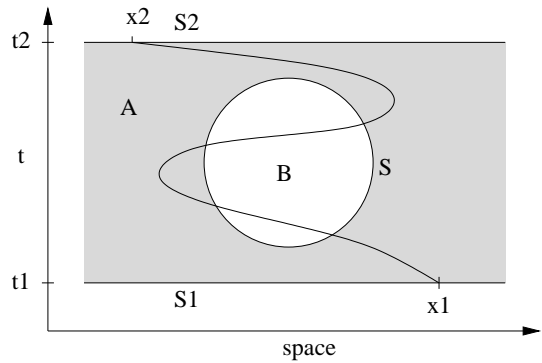}
\caption{A contribution to the path integral over $R=A\cup B$ between
  boundaries $S_1$
  and $S_2$. The inner region $B$ is crossed twice, corresponding to a
  two-particle state on its boundary $S$.}
\label{fig:compose}
\end{figure}

So far everything seems straightforward. However, consider the
following situation. Choose a time interval $[t_1,t_2]$ containing
$B$ and the region of space-time $R$
defined by it. Call its boundaries $S_1, S_2$.
Cut out $B$ from $R$ and call the remainder $A$. Then,
as $R=A\cup B$ the composition rule (T5) requires that $\rho_R$ equals
the composition of $\rho_A$ and $\rho_B$. Say we consider a
one-particle state on $S_1\cup S_2$. Then $\rho_R$ contains an
integral over paths from $S_1$ at $t_1$ to $S_2$ at $t_2$, see
Figure~\ref{fig:compose}. Such a path
may cross the inner region $B$ an arbitrary number of times. However,
we have taken above as the state space $\cH_S$ associated with the
boundary $S$ between $A$ and $B$ the one for one particle. This only
accounts for the paths in the integral that cross $B$ exactly once.
The composition rule seems to be violated. What went wrong?

Looking back at the definition of the one-particle state space $\cH_S$
for $S$ we see that already there is a problem. Namely, the path
integral in the expression for $\rho_B$ only constrains paths at
their starting point and end point. There is no a priori restriction
for them to lie entirely inside $B$. However, we only want to allow to
integrate inside $B$. Thus, how do we deal with paths that would leave
$B$ in between? The answer is rather obvious now. This corresponds to
states with several particles on $B$. We need to let $\cH_S$ be a
direct sum of state spaces \emph{for any number of particles}, i.e.\
$\cH_S=\cH_S^0\oplus \cH_S^1\oplus \cH_S^2\oplus\cdots$ (the
superscript indicates particle number). Now one may restrict the path
integral to paths inside $B$. The occurrence of all paths in a
composition (such as with $A$) is ensured by the summation over all
numbers of particles on $S$. This way we do obtain a consistent
formalism, as guaranteed by the composition rule.

The surprising result is that one needs states with all possible
numbers of particles even in NRQM. Perhaps this is not too surprising
after all. The parallel
to relativistic quantum mechanics is rather apparent. Namely, we
can think of a boost there as tilting a fixed-time boundary (such as
$S_1$). The induced departure from fixed-time boundaries in the
original frame has essentially the effect I have just described.

\begin{figure}
\includegraphics[scale=0.55]{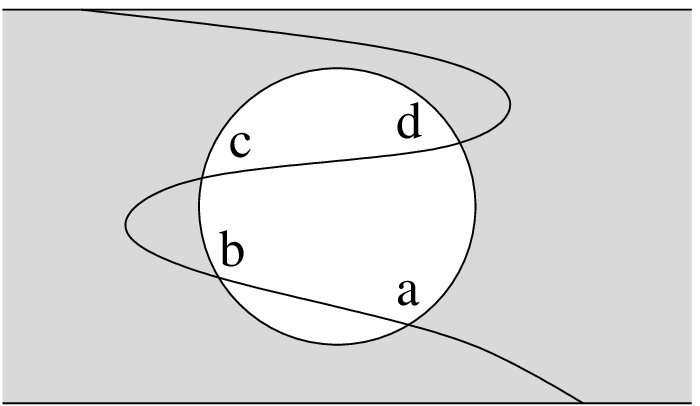}\hspace{0.5cm}
\includegraphics[scale=0.55]{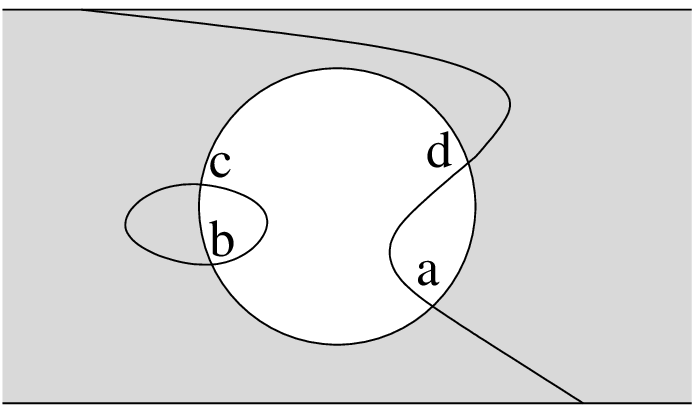}
\caption{Allowing to ``forget'' connectivities introduces pair
  creation and annihilation. The connections $(a,b)$ and $(c,d)$ on
  the left are changed to $(a,d)$ and $(c,b)$ on the right.}
\label{fig:part}
\end{figure}

Note another subtlety. In the configuration spaces for the
multi-particle states, one has to keep track which classical endpoint
is connected to which other one by paths in the integral. If we
remove this restriction we introduce pair creation and annihilation of
particles. Consider Figure~\ref{fig:part}. On the left the
connectivity inside $B$ of the intersection points on the boundary $S$
is described by the pairs $(a,b)$ and $(c,d)$. On the right this is
exchanged to $(a,d)$ and $(c,b)$. The emerging picture might be
interpreted as admitting an extra virtual particle that is created and
subsequently annihilated.

Apart from the fact that we know this to happen in QFT
there is good physical reason to admit this situation also
here. Namely, being an observer on $S$, I see a particle coming into
$B$ at $a$ and $c$ and a particle emerging at $b$ and $d$. If the
particles are identical I have no way to say if it was \emph{the same}
particle that I saw twice or two different particles.
In summary, it turns out that a ``general boundary'' formulation
even of NRQM recovers essential features
of \emph{relativistic} quantum mechanics.

From the point of view of the holographic quantization prescription
the ``accident'' of the failure of the naive (one-particle only)
quantization is also explicable. The starting point of that prescription is
a field theory. NRQM
may be regarded as a field theory, but with the unusual property that
the value of the field $\vx(t)$ is at the same time a coordinate in
space. This resulted then in the problems with the boundaries of the
path integrals that had to be fixed.

\section{Quantum field theory}

Let us move to QFT. Space-time is Minkowski space
and I denote coordinates by $x=(\vx,t)$. I start by considering
regions $R$ determined by time intervals $[t_1,t_2]$.
The encoding of Hilbert spaces and time evolutions in terms of the
TQFT language is as described generically for quantum mechanics
above. The first goal will be to show that the holographic
quantization scheme reduces to the usual Feynman path integral
quantization.

Consider a theory with a massive scalar field $\phi(x)$, for the
moment free. The classical action is thus
$\cS[\phi]=\int \xd x\, \left(\partial_\mu \phi(x) \partial^\mu
\phi(x)- m^2 \phi^2(x)\right)$. 
Considering two time-slices $S_1, S_2$ at $t_1, t_2$, a classical
solution to the field equations in the region $R$ in between is
essentially uniquely determined by 
the values of the field on $S_1$ and $S_2$. Thus, we declare this to be
the configuration space $K_S$ associated with the boundary $S=S_1\cup S_2$
of $R$. It decomposes into $K_S=K_{S_1}\times K_{S_2}$ where each is
the space of field configurations on the respective boundary.
According to (Q1) the state space $\cH_S$ is the space of functions
$C(K_S)$ on $K_S$. It decomposes into $C(K_S)=C(K_{S_1})\tens
C(K_{S_2})$. Consider $\psi_{\vx_1,\dots\vx_n}\in
\cH_{S_1}$ and $\eta_{\vy_1,\dots\vy_n}\in\cH_{S_2}$ given by
$\psi_{\vx_1,\dots\vx_n}(\phi_1)=\phi_1(\vx_1)\cdots \phi_1(\vx_n)$ and
$\eta_{\vy_1,\dots\vy_n}(\phi_2)=\phi_2(\vy_1)\cdots \phi_2(\vy_n)$.
Then, by (Q2) the amplitude of the corresponding state
in $\cH_S$ is given by
\begin{multline}
 \int_{K_{S_1}\times K_{S_2}} \xD \phi_1 \xD \phi_2\,
\phi_1(\vx_1)\cdots \phi_1(\vx_n)\,
 \phi_2(\vy_1)\cdots \phi_2(\vy_n) \\
\int_{\substack{\phi|_{S_1}=\phi_1\\\phi|_{S_2}=\phi_2}}
  \xD\phi\, e^{\frac{i}{\hbar} \cS[\phi]}.
\end{multline}
We recognize this as essentially the transition amplitude between the
``in'' state
$|\phi(\vx_1,t_1)\cdots \phi(\vx_n,t_1)\rangle$ at $t_1$ and the
``out'' state $\langle \phi(\vy_1,t_2)\cdots
\phi(\vy_n,t_2)|$ at $t_2$. We may switch on even a perturbative
interaction in the time interval $[t_1,t_2]$ by modifying the action
in the path integral accordingly.

It turns out that one recovers the complete quantum field
theory correctly. To see this in terms of particle states one has to
repeat the reduction procedure of Lehman, Symanzik and
Zimmermann \cite{LSZ:reduction}, adapted to the situation where
initial and final states are at fixed times $t_1, t_2$ and not in
the infinite past or future. This will be detailed elsewhere. 
An essential role plays the concept of the vacuum as will be explained
in \cite{CDORT:vacuum}.

Having once established the formalism for the special time-slice
boundaries the generalization to arbitrary boundaries is
straightforward following (Q1) and (Q2). Here, no ``accident'' as for
NRQM can happen as the fields have
nothing to do with coordinates in space-time. States on time-slices
can be ``pulled-back'' to any kind of boundary using the composition
rule (T5), as ensured by the form of (Q2). The resulting description
is not only local but also natural in terms of typical
experimental setups.

Consider for example a scattering experiment in
high energy physics. A typical detector has roughly the form of a
sphere with the scattering happening inside (e.g.\ a collision of
incoming beams). The entries for particles and the individual detection
devices are arranged on the surface. At some time $t_1$ the beam is
switched on and at $t_2$ it is switched off. The space-time region $M$
relevant for the experiment is the region inside the sphere times the
time interval $[t_1,t_2]$. The particle inflow and detection happens
on the boundary $S$ of $M$. What seems unusual is that the parts of $S$
that are really interesting and carry the particle states are its
\emph{timelike} components. On the spacelike components at $t_1$ and
$t_2$ there are no particles (we imagine the switching to happen
smoothly). Concerning the interaction term in the Lagrangian it is now
natural to turn it on only inside $M$. Indeed, the particles detected
on the boundary $S$ should (as usual) be thought of as free.

For calculational reasons it will usually be still advantageous to use
particle states that are asymptotic. Indeed, the difference should have
negligible effect on the resulting amplitudes as will be discussed
elsewhere \cite{CDORT:vacuum}. However, there no
longer seems to be a \emph{fundamental} reason to do this. This becomes
rather important for the construction of a non-perturbative theory of
quantum gravity. There, ``asymptotic states'' in terms of a Minkowski
space are not expected to be a useful fundamental concept. The
advantage of a local description is thus crucial.

\section{Quantum gravity}

Both, the ``general boundary'' formulation of quantum theories as well
as the holographic quantization prescription are mainly designed for a
quantum theory of gravity. Then, the background structure for the
TQFT-type
axioms is just that of differentiable manifolds and their
boundaries. Going down to three dimensions it is well known that pure
quantum gravity is a TQFT \cite{Oog:partfunc}, i.e.\ satisfies (T1) -
(T5). What is more, this TQFT is obtainable by
following the quantization prescription given by (Q1) and (Q2).
Using connection variables
the configuration space $K_S$ associated with a
boundary $S$ of a region $M$ is basically the space of flat spin
connections on $S$.
The path integral (Q2) is
rigorously defined through a discretization of $M$ as a spin foam
model.

The role of diffeomorphism invariance is the following here:
If we think of the spin connection as specified by a
connection 1-form $A_\mu(x)$ then $K_S$ is the space of equivalence
classes of such 1-forms under general gauge transformations. These
general gauge transformations are now both, the $SU(2)$ gauge
transformations and diffeomorphisms.

I propose to approach also four-dimensional quantum gravity
using the quantization prescription (Q1) and (Q2). Indeed, the path
integral approach to quantum gravity is well established
\cite{Mis:feyngr,Haw:pathintqg}. The
crucial new ingredient is the admission of arbitrary (in particular
connected) boundaries and  their interpretation
\cite{Oe:catandclock}. A promising context for a non-perturbative
realization of this appear to be spin foam models \cite{Per:sfmodels}
in connection with a renormalization procedure \cite{Oe:renormdisc}.

\acknowledgments

I thank Carlo Rovelli for stimulating discussions and Thomas
Sch\"ucker for a careful reading of the manuscript. This work was
supported through a Marie Curie Fellowship of the European Union.

\bibliography{stdrefs}

\end{document}